\documentclass[aps,prb,reprint]{revtex4-1}

\usepackage{amsmath,amssymb}
\usepackage{bm}
\usepackage{graphicx}

\newcommand{\cD}{\mathcal{D}}
\newcommand{\tD}{\widetilde{\mathcal{D}}}

\begin{document}


\title{Influence of diffuse surface scattering on the stability of
  superconducting phases with spontaneous surface current generated by
  Andreev bound states}


\author{Nobumi Miyawaki}
\author{Seiji Higashitani}
\affiliation{Graduate School of Integrated Arts and Sciences, Hiroshima
University, Kagamiyama 1-7-1, Higashi-Hiroshima 739-8521, Japan}


\date{\today}

\begin{abstract}
  We report a theoretical study on the phase transition between
  superconducting states with and without spontaneous surface current.
  The phase transition takes place due to the formation of surface
  Andreev bound states in unconventional superconductors.  Based on
  the quasiclassical theory of superconductivity, we examine the
  influence of atomic-scale surface roughness on the surface phase
  transition temperature $T_s$. To describe the surface effect, the
  boundary condition for the quasiclassical Green's function is
  parameterized in terms of specularity (the specular reflection
  probability in the normal state at the Fermi level). This boundary
  condition allows systematic study of the surface effect ranging from
  the specular limit to the diffuse limit. We show that diffuse
  quasiparticle scattering at a rough surface causes substantial
  reduction of $T_s$ in the $d$-wave pairing state of high-$T_c$
  cuprate superconductors. We also consider a $p$-wave pairing state
  in which Andreev bound states similar to those in the $d$-wave
  state are generated. In contrast to the $d$-wave case, $T_s$ in the
  $p$-wave state is insensitive to the specularity. This is because
  the Andreev bound states in the $p$-wave superconductor are robust
  against diffuse scattering, as implied from symmetry consideration
  for odd-frequency Cooper pairs induced at the surface; the $p$-wave
  state has odd-frequency pairs with $s$-wave symmetry, while the
  $d$-wave state does not.
\end{abstract}


\maketitle


\section{Introduction}

Theoretical studies of the $d$-wave pairing state in high-$T_c$
cuprate superconductors have predicted a surface state that carries a
spontaneous surface current and locally breaks time-reversal symmetry
$\mathcal{T}$. The authors of Ref.\ \onlinecite{MatsumotoShiba}
demonstrated that a pairing state with $\mathcal{T}$-breaking symmetry
such as $d+is$ is stabilized near the surface by a subdominant pairing
interaction and this surface state with broken $\mathcal{T}$ generates
a spontaneous current.  The spontaneous surface current was later
shown to occur also in the absence of subdominant interactions
\cite{SeijiJPSJ1997}.  The origin of the local symmetry breaking lies
in the existence of Andreev bound states (ABSs) that form, in the
presence of $\mathcal{T}$, a flat band at zero energy (Fermi level)
\cite{Hu,TanakaKashiwaya,KashiwayaTanaka}. Those midgap ABSs drive the
instability of the $\mathcal{T}$-preserving $d$-wave phase toward a
$\mathcal{T}$-breaking phase. In the latter superconducting (SC)
phase, the bound-state band is shifted from the Fermi level and
thereby the surface free energy can be lowered \cite{SigristPTP}.  The
self-induced vector potential associated with the spontaneous current
provides a mechanism for the energy shift \cite{SeijiJPSJ1997,
  Lofwander}.  The subdominant order parameter itself also brings
about the energy shift \cite{MatsumotoShiba, Fogelstrom}. In
restricted geometries such as thin films \cite{VorontsovPRL,
  SeijiJPSJ2015, MiyawakiPRB2015, MiyawakiJLTP2017}, a direct phase
transition from the normal state to the $\mathcal{T}$-breaking state
was shown to be possible when the confinement size is of the order of
the coherent length $\xi_0$. Recently, spontaneous generation of a
vortex chain structure was predicted to occur along the surface of the
cuprate superconductors \cite{Hakansson, Holmvall, HolmvallPHD}.

\begin{figure}[t]
  \includegraphics[scale=1.1]{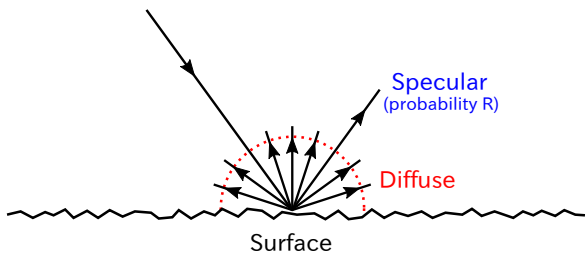}
  \caption{\label{fig:rough_surface} Scattering at a rough surface
    parameterized by specularity (specular reflection probability)
    $R$.  An incident quasiparticle in the normal state at the Fermi
    level is scattered specularly with probability $R$ and diffusively
    with probability $1-R$ [see Eq.\ \eqref{eq:S2-R}].  The specular
    and diffuse limits correspond to $R = 1$ and $R = 0$,
    respectively.}
\end{figure}

In this paper, we are concerned with the surface phase transition
between the SC states with and without the spontaneous surface
current.  In general, the surface physics sensitively depends on the
nature of the boundary condition.  For example, surface roughness
causes significant modification of the surface density of states
(SDOS) in superconductors and superfluids \cite{Zhang, Yamada,
  Yamamoto, VorontsovSauls, NagaiJPSJ, MurakawaPRL, MurakawaJPSJ,
  Okuda}. In the case of the $d$-wave SC state, diffuse quasiparticle
scattering by the surface roughness results in substantial broadening of
SDOS at zero energy \cite{Yamada}.  The broadening of zero-energy SDOS
suggests the reduction of the surface phase transition temperature
$T_s$ \cite{Barash}.  Here, we address the rough surface problem with
the purpose of evaluating the robustness of the $\mathcal{T}$-breaking
SC phase against diffuse surface scattering.  We parameterize the
boundary problem in terms of the specularity of the surface
\cite{NagaiJPSJ, MurakawaPRL, MurakawaJPSJ, Okuda}. This
parameterization allows us to treat the surface effect ranging from
the specular limit to the diffuse limit in a unified way (Fig.\
\ref{fig:rough_surface}).  For simplicity, we do not take into
consideration impurity effects \cite{Barash}, subdominant pairing
channels \cite{MatsumotoShiba, Fogelstrom}, and the possibility of the
surface vortex chain state \cite{Hakansson, Holmvall, HolmvallPHD}.

We consider not only the $d$-wave state but also a $p$-wave (polar)
state (Fig.\ \ref{fig:system}). The two SC states have a common
symmetry such that the gap function felt by quasiparticles changes
sign for specular reflection processes. Because of this symmetry,
the midgap ABSs appear in both superconductors \cite{Hu,
  HaraNagai, OhashiTakada}. When the surface is specular, the midgap
ABSs manifest in SDOS as a zero-energy peak.  As mentioned above, this
peak in the $d$-wave state is broadened in the presence of surface
roughness.  On the other hand, SDOS in the $p$-wave polar state is
hardly affected by diffuse scattering \cite{Yamamoto}.  We show that
$T_s$ in the $p$-wave state is insensitive to surface roughness, while
in the $d$-wave state the broadening of zero-energy SDOS gives rise to
a substantial reduction of $T_s$.  The difference between the two SC
states in the sensitivity to surface roughness can qualitatively be
understood from symmetry consideration for odd-frequency Cooper pairs
induced at the surface of the two SC states. This point will be
discussed in the final part of Sec.\ \ref{sec:results}.

Our calculations are based on the quasiclassical theory of
superconductivity \cite{Eilenberger,LarkinOvchinnikov}. We outline the
theoretical formulation in Sec.\ \ref{sec:formulation}.  The rough
surface effect is described by random $S$-matrix theory
\cite{NagatoJLTP}, from which one can obtain the specularity-dependent
boundary condition for the quasiclassical equation.  We numerically
solve Maxwell's equations along with the quasiclassical equation to
determine the vector potential spontaneously induced in the
$\mathcal{T}$-breaking SC phase. The surface value of the vector
potential, which is proportional to the total spontaneous magnetic
field, exhibits a temperature dependence typical for a second-order
phase transition. We determine the transition temperature $T_s$ for
various values of specularity by calculating the linear response of
the system to the vector potential. Those numerical results are
presented in Sec.\ \ref{sec:results}.  Our conclusions are summarized in
Sec.\ \ref{sec:conclusion}.

\section{Quasiclassical theory}
\label{sec:formulation}

The quasiclassical theory is formulated in terms of a Green's function
$\hat{g}(\bm{r}, \hat{p}, \epsilon)$, which is a $4 \times 4$ matrix
in Nambu space.  Here, $\bm{r}$ is the real-space position vector,
$\hat{p}$ a unit vector to specify the Fermi-surface position, and
$\epsilon$ a complex energy variable.  The four-dimensional Nambu
space is spanned by spin and particle-hole degrees of freedom.  From
symmetry consideration, the quasiclassical Green's function $\hat{g}$
is found to have the matrix structure (Appendix \ref{appendix:A})
\begin{gather}
  \hat{g}(\bm{r}, \hat{p}, \epsilon) =
  \begin{bmatrix}
    ig(\bm{r}, \hat{p}, \epsilon) &
    f(\bm{r}, \hat{p}, \epsilon) \\
    f(\bm{r}, -\hat{p}, -\epsilon^*)^* &
    -ig(\bm{r}, -\hat{p}, -\epsilon^*)^* \\
  \end{bmatrix},
  \label{eq:g-mat-ex}
\end{gather}
where the elements are $2 \times 2$ matrices in spin space. The spatial
evolution of $\hat{g}$ is governed by the Eilenberger equation
\begin{equation}
  i\hbar \bm{v}_{\hat{p}} \cdot \nabla_{\bm{r}} \hat{g}
  = [\hat{g},\ (\epsilon - \hat{\Delta})\hat{\rho}_3]
  \label{eq:Eileneq}
\end{equation}
supplemented by the normalization condition 
\begin{equation}
  \hat{g}^2(\bm{r}, \hat{p}, \epsilon) = -1 
\end{equation}
and appropriate boundary conditions depending on the geometry of
system.  The gradient term on the left-hand side of Eq.\
\eqref{eq:Eileneq} connects $\hat{g}$ at different spatial points on a
straight line corresponding to the classical trajectory along the
Fermi velocity $\bm{v}_{\hat{p}}$. On the right-hand side,
\begin{equation}
  \hat{\Delta} =
  \begin{bmatrix}
    0 && \Delta(\bm{r}, \hat{p}) \\ \Delta(\bm{r}, \hat{p})^\dag && 0 \\
  \end{bmatrix}
\end{equation}
is the Nambu-space gap matrix and $\hat{\rho}_3$ is the third Pauli
matrix in particle-hole space.  In superconductors with a spontaneous
surface current, a magnetic field $\bm{B}(\bm{r}) = \nabla_{\bm r}
\times \bm{A}(\bm{r})$ is induced near the surface.  The
current-carrying state can be treated by replacing $\epsilon$ in Eq.\
\eqref{eq:Eileneq} as
\begin{equation}
  \epsilon \to 
  \epsilon - \hbar \bm{v}_{\hat{p}}\cdot\bm{Q}(\bm{r})/2,
  \label{eq:e-to-e}
\end{equation}
where $\bm{Q}(\bm{r}) = -(2e/c\hbar)\bm{A}(\bm{r})$ with $e$ ($e<0$)
being the electron charge and $c$ the speed of light.  The magnetic
field $\bm{B}(\bm{r})$ is related to the current density
$\bm{J}(\bm{r})$ by Maxwell's equation
\begin{equation}
  \nabla_{\bm r} \times \bm{B}(\bm{r}) = \frac{4\pi}{c}\bm{J}(\bm{r}).
\end{equation}

\begin{figure}[t]
  \includegraphics[scale=1.2]{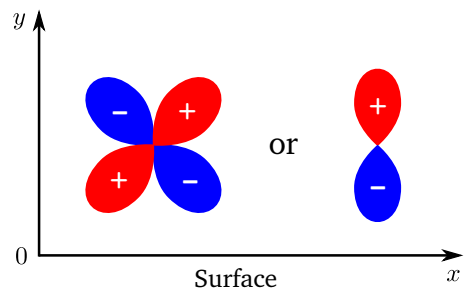}
  \caption{\label{fig:system} Semi-infinite superconductor.  The
    surface is located at $y = 0$ and a quasi-two-dimensional
    superconductor occupies the $y > 0$ space. The superconductor is
    in the $d_{xy}$-wave (left) or $p_{y}$-wave (right) pairing states.}
\end{figure}

The gap matrix $\Delta(\bm{r}, \hat{p})$ and the current density
$\bm{J}(\bm{r})$ can be determined from $\hat{g}(\bm{r}, \hat{p},
\epsilon)$ on the imaginary axis of the complex $\epsilon$ plane,
i.e., $\hat{g}(\bm{r}, \hat{p}, i\epsilon_n)$ at the Matsubara
energies $\epsilon_n = (2n+1)\pi/\beta$ with $n = 0, \pm 1, \pm 2,
\cdots$, and $\beta = 1/k_BT$ being the inverse temperature.  The
corresponding equations are
\begin{align}
  &\Delta(\bm{r}, \hat{p}) =
  N(0)\frac{\pi}{\beta} \sideset{}{'}\sum_{\epsilon_n}
  \left\langle
    V_{\hat{p}\hat{p}'} f(\bm{r}, \hat{p}', i\epsilon_n)
  \right\rangle_{\hat{p}'},
  \label{eq:gapeq}
  \\
  &\bm{J}(\bm{r}) =
  e N(0) \frac{\pi}{\beta}\sum_{\epsilon_n}
  {\rm Im} \left\langle 
    \bm{v}_{\hat{p}}\, {\rm Tr}\,g(\bm{r}, \hat{p}, i\epsilon_n)
  \right\rangle_{\hat{p}},
  \label{eq:J-g}
\end{align}
where $N(0)$ is the density of states (per spin) in the normal state
at the Fermi level and $V_{\hat{p}\hat{p}'}$ the pairing
interaction.  The notation
\begin{equation}
  \langle \cdots \rangle_{\hat{p}}
  \equiv
  \frac{\int (\cdots) {d^2 p_F}/{|\bm{v}_{\hat{p}}|}}
  {\int {d^2 p_F}/{|\bm{v}_{\hat{p}}|}}
\end{equation}
denotes the average over the Fermi surface.  The prime on the sum in
Eq.\ \eqref{eq:gapeq} means that a cutoff is necessary for the
Matsubara sum.

From $\hat{g}$, one can also get information on the quasiparticle
density of states.  The angle-resolved local density of states,
normalized to be unity at an energy $E$ sufficiently larger than the SC
gap, is given in terms of the diagonal elements of $\hat{g}$ with
$\epsilon$ on the real axis:
\begin{equation}
  n(\bm{r}, \hat{p}, E) =
  {\rm Re}\left[\frac{1}{2} {\rm Tr}\,g(\bm{r}, \hat{p}, E + i\delta)\right],
  \label{eq:dos-g}
\end{equation}
where $\delta$ is an infinitesimal positive constant defining the
retarded Green's function.

In the actual calculation of the quasiclassical Green's function, we used
the Riccati parameterization method \cite{EschrigPRB}. In this method,
the spin-space matrix Green's functions $g$ and $f$ are expressed as
(Appendix \ref{appendix:A})
\begin{align}
  g(\bm{r}, \hat{p}, \epsilon)
  &= \frac{2}
  {1 - \cD(\bm{r}, \hat{p}, \epsilon)\cD(\bm{r}, -\hat{p}, -\epsilon^*)^*} - 1,
  \\
  f(\bm{r}, \hat{p}, \epsilon)
  &= [g(\bm{r}, \hat{p}, \epsilon) + 1]\cD(\bm{r}, \hat{p}, \epsilon),
\end{align}
with $\cD(\bm{r}, \hat{p}, \epsilon)$ obeying the Riccati-type
differential equation
\begin{align}
  \hbar \bm{v}_{\hat{p}}\cdot\nabla_{\bm r} \cD = 2i\epsilon \cD +
  \Delta(\bm{r}, \hat{p}) - \cD \Delta(\bm{r},\hat{p})^\dag \cD.
  \label{eq:RCTeq}
\end{align}
We note again that $\epsilon$ in Eq.\ \eqref{eq:RCTeq} is replaced by
Eq.\ \eqref{eq:e-to-e} when surface current flows.

We apply the quasiclassical theory to a semi-infinite geometry as
depicted in Fig.\ \ref{fig:system}. A quasi-two-dimensional
superconductor with a flat surface at $y = 0$ occupies the $y > 0$
space.  The quasi-two-dimensionality is described by a cylindrical
Fermi surface with an isotropic Fermi velocity $|\bm{v}_{\hat{p}}| =
v_F$. The surface may have atomic-scale irregularity, though it is
assumed to be macroscopically flat. We consider the effect of
the surface roughness by parameterizing the boundary condition for
Eq.\ \eqref{eq:RCTeq} in terms of the specularity $R$ defined as the
specular reflection probability in the normal state at the Fermi level
(Fig.\ \ref{fig:rough_surface}).  The boundary condition is obtained
from the random-$S$ matrix theory developed in Ref.\
\onlinecite{NagatoJLTP}. The outline of this theory and the explicit
expression for the boundary condition are given in Appendix
\ref{appendix:B}. We characterize the SC phase with broken
$\mathcal{T}$ by the vector fields
\begin{equation*}
  \bm{Q}(\bm{r}) = Q_x(y) \bm{e}_x,\ 
  \bm{B}(\bm{r}) = B_z(y) \bm{e}_z,\ 
  \bm{J}(\bm{r}) = J_x(y) \bm{e}_x,
\end{equation*}
where $\bm{e}_i$ is the unit vectors along the $i$-axis of real-space
coordinate.

The above SC system is assumed to be in $d_{xy}$-wave or $p_y$-wave
states with the gap matrix
\begin{align}
  \Delta(\bm{r}, \hat{p}) = \Delta_0(y)\zeta(\hat{p}) s_\sigma.
  \label{eq:Delta-zeta}
\end{align}
For the $d_{xy}$-wave state, $\zeta(\hat{p}) =
2\sqrt{2}\,\hat{p}_x\hat{p}_y$ and $s_\sigma = i\sigma_2$. For the
$p_y$-wave state, $\zeta(\hat{p}) = \sqrt{2}\,\hat{p}_y$ and $s_\sigma
= \bm{s}\cdot\bm{\sigma}i\sigma_2$. Here, $\bm{\sigma} = (\sigma_1,
\sigma_2, \sigma_3)$ is the Pauli matrix and $\bm{s}$ is a unit vector
in spin space. Because our model system has rotational symmetry in spin
space, the direction of $\bm{s}$ may be chosen arbitrarily. The basis
function $\zeta(\hat{p})$ is normalized as $\langle \zeta^2(\hat{p})
\rangle_{\hat{p}} = 1$.  The single-component SC states can be
characterized by the pairing interaction of the form
$V_{\hat{p}\hat{p}'} = V\zeta(\hat{p})\zeta(\hat{p}')$. The
interaction parameter $V$ is related to the transition temperature
$T_c$ between the normal and bulk-SC states by
\begin{equation}
  \frac{1}{N(0)V} = 2\pi k_B T_c
  \sideset{}{'}\sum_{\epsilon_n > 0} \frac{1}{\epsilon_n}
  \approx \ln (1.13\epsilon_c/k_BT_c),
\end{equation}
where $\epsilon_c$ denotes the cutoff energy for the Matsubara sum.

\section{Numerical results}
\label{sec:results}

For numerical calculation of $Q_x(y)$, $B_z(y)$, and $J_x(y)$, we
introduce the dimensionless quantities
\begin{align}
  q_x(y) &= \xi_0 Q_x(y),\\
  b_z(y) &= \frac{2|e|}{\hbar c}\lambda_0\xi_0 B_z(y),\\
  j_x(y) &= \frac{8\pi|e|}{\hbar c^2}\lambda_0^2\xi_0 J_x(y)
  = \frac{J_x(y)}{\pi |e| v_F N(0) k_B T_c},
\end{align}
where $\xi_0 = \hbar v_F / 2 \pi k_B T_c$ is the coherence length and
$\lambda_0 = (c^2/4\pi e^2N(0) v_F^2)^{1/2}$ is the London penetration
depth at $T = 0$. The dimensionless fields are determined from
Maxwell's equations
\begin{gather}
  \lambda_0 \frac{dq_x(y)}{dy} = -b_z(y),
  \label{eq:Meq1}
  \\
  \lambda_0 \frac{db_z(y)}{dy} = j_x(y),
  \label{eq:Meq2}
\end{gather}
along with $j_x(y)$ obtained from Eq.\ \eqref{eq:J-g}.  The boundary
conditions are $q_x(\infty) = 0$ and $b_z(0) = 0$. In the
self-consistent calculation of the fields, we neglect, for simplicity,
the surface pairbreaking effect and put $\Delta_0(y) =
\Delta_0(\infty)$. This approximation will not be serious because the
low-energy structure of SDOS is insensitive to the self-consistency of
the gap function \cite{NagatoNagaiPRB}.

\begin{figure}[t]
  \includegraphics[scale=0.9]{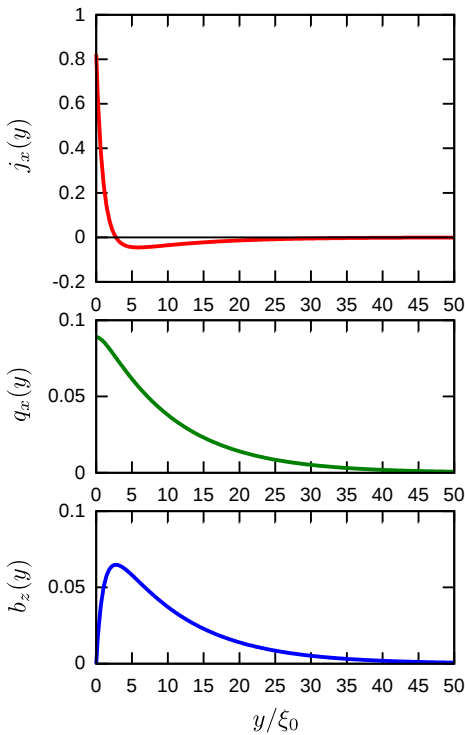}
  \caption{\label{fig:dxy_fields_specular} Spatial distribution of
    $j_x(y)$, $q_x(y)$, and $b_z(y)$ in the $d_{xy}$ superconductor
    with $\lambda_0/\xi_0 = 10.0$ at $T/T_c = 0.02$. The surface is
    assumed to be specular ($R = 1$).}
\end{figure}

In Fig.\ \ref{fig:dxy_fields_specular}, we plot the typical spatial
distribution of the fields $j_x(y)$, $q_x(y)$, and $b_z(y)$ induced
spontaneously in the $d_{xy}$ superconductor. The results are shown
for $R = 1$.  The current $j_x(y)$ takes a large positive value at the
surface ($y = 0$) owing to the formation of midgap ABSs. As the
distance $y$ from the surface increases, $j_x(y)$ decreases and
becomes negative at $y \sim \xi_0$. The negative (screening) current
prevents the spontaneous magnetic field $b_z(y)$ from penetrating into
the superconductor. The total current $\int_0^\infty dy\, j_x(y)$
vanishes \cite{SeijiJPSJ1997,OhashiMomoi,KusamaOhashi}, as assured by
Maxwell's equation \eqref{eq:Meq2} with the boundary condition
$b_z(0) = 0$.  The fields for $ R \neq 1$ exhibit similar $y$
dependence.

\begin{figure}[t]
  \includegraphics[scale=0.9]{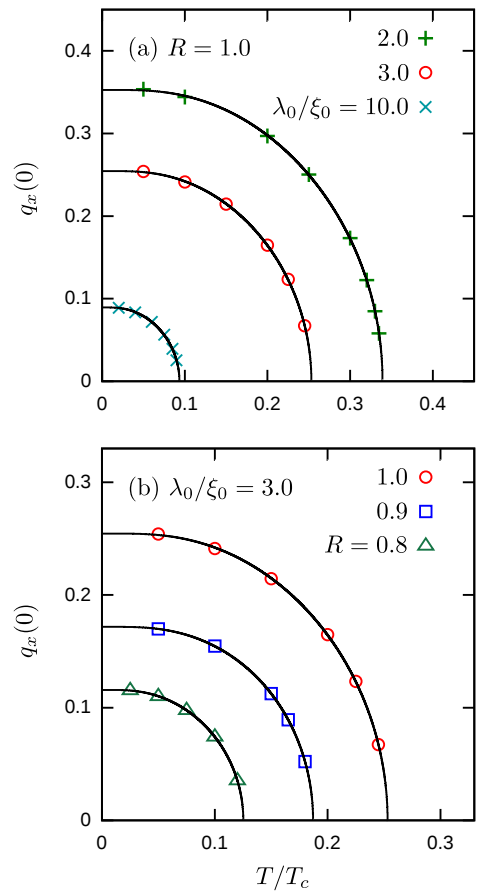}
  \caption{\label{fig:dxy_q_t} Temperature dependence of $q_x(0)$ in
    the $d_{xy}$ superconductor for (a) several $\lambda_0/\xi_0$ at
    $R=1.0$ and (b) several $R$ at $\lambda_0/\xi_0 = 3.0$.  The
    symbols are the numerical results and the solid lines are fits
    using Eq.\ \eqref{eq:fit_qx0}.}
\end{figure}

\begin{figure}[h]
  \includegraphics[scale=0.9]{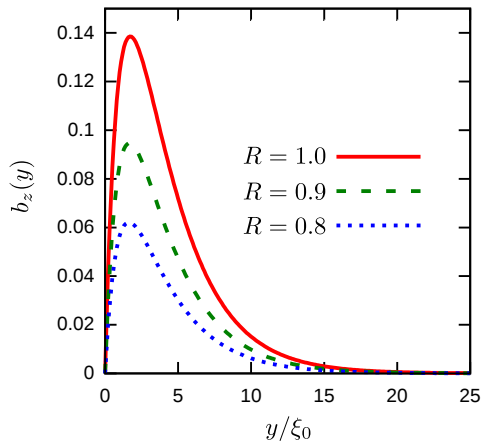}
  \caption{\label{fig:dxy_bz_rough_surface} Spontaneous magnetic field
    $b_z(y)$ in the $d_{xy}$ superconductor with $\lambda_0/\xi_0 =
    3.0$ at $T / T_c = 0.05$.}
\end{figure}

The spontaneous surface current appears at low temperatures after a
second-order phase transition from the conventional
$\mathcal{T}$-preserving SC state. To demonstrate the surface phase
transition in the $d_{xy}$ superconductor, we plot in Fig.\
\ref{fig:dxy_q_t} the temperature dependence of $q_x(0)$, which is
proportional to the total magnetic field induced by the spontaneous
current [see Eq.\ \eqref{eq:Meq1}].  The symbols are the results
obtained by numerically solving Maxwell's equations at several
temperatures.  The solid lines are fits using
\begin{equation}
  q_x(0) = C_1 \tanh\left(C_2 \sqrt{C_3/t -1} \right),
  \label{eq:fit_qx0}
\end{equation}
where the $C_i$'s are fitting parameters and $t = T/T_c$ is the reduced
temperature.  The numerical data are well fitted by Eq.\
\eqref{eq:fit_qx0}, in which a second-order phase transition is
assumed to take place at $t = C_3$ corresponding to the surface phase
transition temperature $T_s$ scaled by $T_c$.  As we increase the
parameter $\lambda_0 / \xi_0$, the reduced transition temperature $T_s
/ T_c$ decreases [Fig.\ \ref{fig:dxy_q_t} (a)].  The origin of this
property is the different length scales between the surface current
carried by ABSs and the conventional screening current. The former is
localized within the surface region of width $\sim \xi_0$. The latter
flows within a width $\sim \lambda_0$. To satisfy the condition of
vanishing total current at a finite $q_x(0)$, larger ABS current and
therefore lower temperature is required for larger $\lambda_0/\xi_0$.
Figure \ref{fig:dxy_q_t} (b) demonstrates the effect of diffuse
surface scattering on $q_x(0)$. The reduced transition temperature
$T_s/T_c$ is suppressed by diffuse scattering and depends rather
sensitively on the specularity $R$.  The corresponding suppression of
$b_z(y)$ at $T/T_c = 0.05$ is shown in Fig.\
\ref{fig:dxy_bz_rough_surface}.

To study the rough surface effect on $T_s/T_c$ in more detail, we
solved the linearized Maxwell's equations numerically
\begin{equation}
  \int_0^\infty dy' K(y,y')\, q_x(y')
  = \lambda_0^2 \frac{d^2q_x(y)}{dy^2}.
  \label{eq:linear_Meq}
\end{equation}
The left-hand side corresponds to the linear response of $-j_x(y)$ to
$q_x(y)$. The kernel $K(y,y')$ can be obtained by expanding the
quasiclassical Green's function $g$ to linear order in $q_x(y)$ and
substituting the linear deviation into Eq.\ \eqref{eq:J-g}. The
resulting explicit formula is so lengthy that it is not shown here.
We note only that $K(y,y')$ is real and symmetric under the exchange
of $y$ and $y'$.

To solve Eq.\ \eqref{eq:linear_Meq}, we used the finite difference
formulas
\begin{equation}
  \frac{dq_x}{dy} = \frac{q_{i+1} - q_{i-1}}{2h},\ \
  \frac{d^2q_x}{dy^2} = \frac{q_{i+1} - 2q_i + q_{i-1}}{h^2},
\end{equation}
where $q_i = q_x(ih)$ with $i$ being an integer. Evaluating the $y'$
integral in Eq.\ \eqref{eq:linear_Meq} using the trapezoidal rule, we
obtain
\begin{gather*}
  \frac{K_{i0} q_0 + K_{iN} q_N}{2} + \sum_{j=1}^{N-1} K_{ij} q_j
  = \mu (q_{i+1} - 2q_i + q_{i-1}),
  \\
  b_z(0) \propto (q_{1} - q_{-1}) / 2h = 0,\ \ q_N = 0,
\end{gather*}
where $K_{ij} = h K(ih, jh)$ and $\mu = \lambda_0^2 / h^2$. This set
of equations can be cast into the form of the generalized eigenvalue
equation $\mathbb{A}\vec{q} = \mu \mathbb{B} \vec{q}$ with
$\mathbb{A}$ being a real symmetric matrix and $\mathbb{B}$ being a
positive-definite real symmetric matrix.  Using the GNU Scientific
Library, we solved it to obtain the eigenvalue $\mu$ at a given $t =
T/T_c$ (we performed the calculations down to $t = 0.01$).  The
resulting $\mu$-$t$ relation gives $T_s/T_c$ as a function of
$\lambda_0/\xi_0$. From the numerical calculation, we found that the
maximum eigenvalue $\mu_{\rm max}$ reproduces $T_s/T_c$ determined
from the full (nonlinear) Maxwell's equation (Fig.\ \ref{fig:dxy_q_t}).

\begin{figure}[t]
  \includegraphics[scale=1.2]{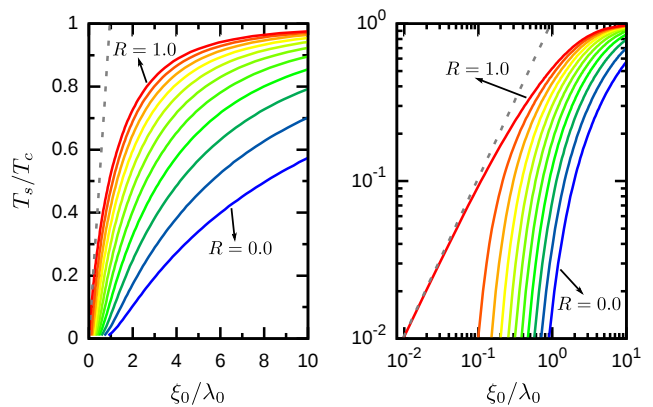}
  \caption{\label{fig:dxy_t_lmd} Reduced transition temperature
    $T_s/T_c$ in the $d_{xy}$ superconductor as a function of
    $\xi_0/\lambda_0$. The left panel is the linear plot of $T_s/T_c$
    vs $\xi_0/\lambda_0$ and the right panel the corresponding log-log plot.  The
    solid lines are, from right to left, the numerical
    results obtained by changing specularity $R$ from zero to unity in
    increments of 0.1.  The dashed line corresponds to Eq.\
    \eqref{eq:Barash}.}
\end{figure}

\begin{figure}[t]
  \includegraphics[scale=1.2]{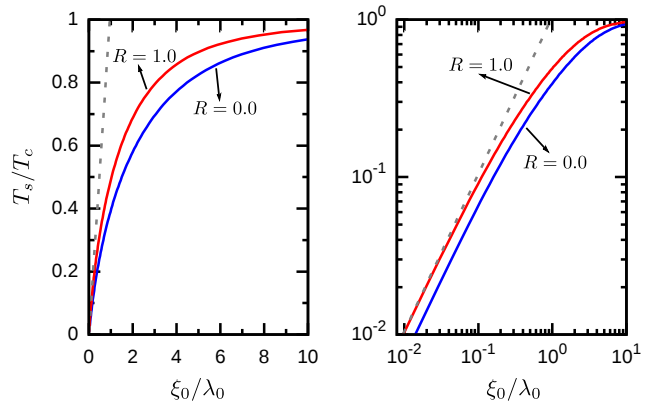}
  \caption{\label{fig:py_t_lmd} Reduced transition temperature
    $T_s/T_c$ in the $p_{y}$ superconductor as a function of
    $\xi_0/\lambda_0$. The left panel is the linear plot of $T_s/T_c$
    vs $\xi_0/\lambda_0$ and the right panel is its log-log plot.  The
    solid lines are the numerical results for $R = 0.0$ and $1.0$. The
    dashed line corresponds to Eq.\ \eqref{eq:Barash}.}
\end{figure}

\begin{figure}[t]
  \includegraphics[scale=1.2]{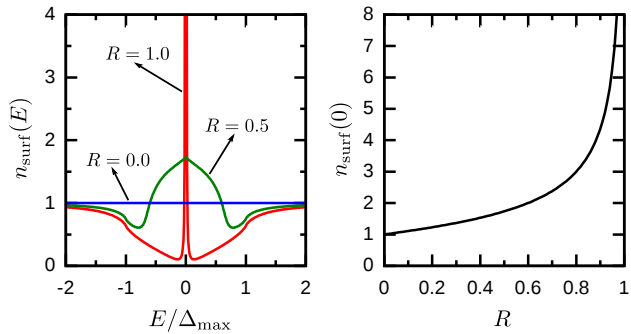}
  \caption{\label{fig:dxy_sdos_q_zero} SDOS in the $d_{xy}$-wave SC
    phase without spontaneous surface current. The left panel shows
    the energy dependence of SDOS for $R = 0.0$, $0.5$, and $1.0$.  In
    calculating these results, we choose $\delta$ in Eq.\
    \eqref{eq:dos-g} to be $10^{-3}\Delta_{\rm max}$, where
    $\Delta_{\rm max} = \sqrt{2}\Delta_0$ is the maximum value of the
    $\hat{p}$-dependent energy gap in the bulk SC state. In the right
    panel, SDOS at zero energy, Eq.\ \eqref{eq:n_surf_0_ex}, is
    plotted as a function of $R$.}
\end{figure}

In Fig.\ \ref{fig:dxy_t_lmd}, we plot $T_s/T_c$ in the $d_{xy}$-wave
state as a function of $\xi_0/\lambda_0$.  The same plot for the
$p_y$-wave state is shown in Fig.\ \ref{fig:py_t_lmd}.  The solid
lines are the numerical results for various values of $R$. The dashed
line represents the approximate formula \cite{Barash}
\begin{equation}
  \frac{T_s}{T_c} = \frac{\pi}{3} \frac{\xi_0}{\lambda_0},
  \label{eq:Barash}
\end{equation}
which can be applied to strong type-II $d_{xy}$-wave and $p_{y}$-wave
superconductors with $R = 1.0$. When $R = 1.0$, the two
superconductors have almost the same $T_s/T_c$. However, the rough
surface effect on $T_s/T_c$ is quite different between the two
states. Diffuse surface scattering results in a substantial reduction of
$T_s/T_c$ in the $d_{xy}$-wave case. On the other hand, $T_s / T_c$ in
the $p_y$-wave state is insensitive to surface roughness.  This marked
difference can be understood qualitatively by observing SDOS in the
absence of surface current.  In Fig.\ \ref{fig:dxy_sdos_q_zero}, we
plot the total SDOS, the surface value $n_{\rm surf}(E)$ of $\langle
n(\bm{r}, \hat{p}, E) \rangle_{\hat{p}}$, in the $d_{xy}$
superconductor. In the specular limit, there is a delta-function peak
at zero energy originating from midgap ABSs.  This peak is broadened
by diffuse scattering and the midgap SDOS, $n_{\rm surf}(0)$,
decreases steeply as the specularity $R$ decreases from unity.  We can
show that $n_{\rm surf}(0)$ in the $d_{xy}$ superconductor depends on
$R$ as \cite{Yamada}
\begin{equation}
  n_{\rm surf}(0) = \frac{1}{2}
  \left(
    \frac{1+\sqrt{R}}{\sqrt{1 - \sqrt{R}}} +
    \frac{\sqrt{1 - \sqrt{R}}}{1+\sqrt{R}}
  \right).
  \label{eq:n_surf_0_ex}
\end{equation}
In the diffuse limit, $n_{\rm surf}(0)$ is suppressed to unity (then
SDOS in the whole energy region coincides with that of the normal
state \cite{Yamada}).  The broadening of the midgap SDOS implies the
reduction of the ABS current, resulting in the decrease of $T_s/T_c$
with $R$. In the $p_y$-wave state, SDOS also has a zero-energy peak.
In contrast to the $d_{xy}$ case, however, SDOS in the $p_y$-wave
state is quite robust against diffuse scattering \cite{Yamamoto}.

The robustness of the midgap SDOS is closely related to the symmetry
of odd-frequency Cooper pairing. As has been shown in the studies of
boundary effects in superconductors and superfluids, ABSs appear
accompanied by odd-frequency pairs (for a review, see Ref.\
\onlinecite{TanakaJPSJ2012}). Moreover, there is a relationship
between the midgap density of states and the odd-frequency pair
amplitude, which states the equivalence between them
\cite{SeijiPRB2012, TsutsumiMachida, SeijiPRB2014, MizushimaOddFreq}.
Fermi statistics requires that the odd-frequency pairs in spin-singlet
and spin-triplet states have odd-parity and even-parity symmetries,
respectively. The robustness of the midgap SDOS in the $p_y$-wave
superconductor is supported by the triplet odd-frequency $s$-wave
pairing induced at the surface.

\section{Conclusion}
\label{sec:conclusion}

We have examined numerically the influence of surface roughness on the
instability temperature $T_s$ toward the appearance of a spontaneous
surface current in unconventional superconductors. This surface phase
transition is driven by midgap Andreev bound states such as formed in
the $d$-wave pairing state of high-$T_c$ cuprate superconductors
\cite{SeijiJPSJ1997, Lofwander, Barash}. Considering strong type-II
superconductors like the cuprates and assuming the surface to be
specular, one can analytically estimate $T_s$ and obtain the result
$T_s \sim (\xi_0 / \lambda_0) T_c$ \cite{Lofwander, Barash}. Our
numerical calculation for the specular surface reproduces this
result well. In actual systems, the surface inevitably has atomic-scale
surface roughness giving rise to diffuse scattering of
quasiparticles. In our theory, the rough surface effect is
parameterized in terms of the surface specularity (Fig.\
\ref{fig:rough_surface}). We have calculated the specularity
dependence of $T_s/T_c$ in the $d$-wave superconductor and found that
the broadening of the midgap Andreev bound states at a rough surface
causes substantial reduction of $T_s / T_c$ even for such a large
specularity as $0.9$ (Fig.\ \ref{fig:dxy_t_lmd}).

We have compared the result of $T_s / T_c$ for the $d$-wave state to
that for the $p$-wave polar state in which the gap function has a
momentum-direction dependence \cite{Hu, HaraNagai, OhashiTakada}
responsible for the generation of the midgap Andreev bound states,
similar to those in the $d$-wave superconductor (Fig.\
\ref{fig:system}). For the $p$-wave superconductor, we found that
$T_s/T_c$ is insensitive to specularity (Fig.\
\ref{fig:py_t_lmd}). This difference from the $d$-wave
case can be accounted for by the fact that in the $p$-wave state there
exist odd-frequency $s$-wave Cooper pairs behind the midgap states.
The presence of the odd-frequency $s$-wave pairs assures that the
midgap states are robust against diffuse surface scattering.

In the present work, we have assumed that the spontaneous surface
current $\bm{J}(\bm{r})$ depends only on the coordinate perpendicular
to the surface.  This assumption excludes the possibility of a
spontaneously-induced vortex chain structure, which has recently been
predicted to appear along the surface of the high-$T_c$ cuprates
\cite{Hakansson, Holmvall, HolmvallPHD}.  The surface phase transition
temperature to the vortex chain state was reported to be higher
than that for the surface state considered here. It should be noted,
however, that the theoretical analysis is based on the specular
surface model. The rough surface effect on the stability of this novel
surface state is an important issue that remains to be examined.

\begin{acknowledgments}
  We thank M. Ashida for valuable advice on the numerical method for
  calculating the results in Sec.\ \ref{sec:results}.  We also thank
  Y. Nagato and K. Nagai for helpful discussions about the rough
  surface effects on the Andreev bound states. This work was supported
  in part by the JSPS KAKENHI Grant Number 15K05172.
\end{acknowledgments}

\appendix

\section{Symmetry and Nambu-space matrix structure of 
the quasiclassical Green's function}
\label{appendix:A}

The quasiclassical Green's function $\hat{g}(\bm{r}, \hat{p},
\epsilon)$ defined as a $4 \times 4$ Nambu-space matrix has the symmetry
\cite{SereneRainer}
\begin{align}
  \hat{g}(\bm{r}, \hat{p}, \epsilon)
  &= \hat{\rho}_1 \widetilde{\hat{g}}(\bm{r}, \hat{p}, \epsilon) \hat{\rho}_1
  \label{eq:symrel-g-1}\\
  &= \hat{\rho}_3 \hat{g}(\bm{r}, \hat{p}, \epsilon^*)^\dag \hat{\rho}_3,
  \label{eq:symrel-g-2}
\end{align}
where $\hat{\rho}_i$'s are the Pauli matrices in particle-hole space
and the tilde transform in Eq.\ \eqref{eq:symrel-g-1} is defined as
\begin{equation}
  \widetilde{X}(\bm{r}, \hat{p}, \epsilon)
  = X(\bm{r}, -\hat{p}, -\epsilon^*)^*.
\end{equation}
It follows from Eq.\ \eqref{eq:symrel-g-1} that $\hat{g}$ has the
matrix structure
\begin{equation}
  \hat{g}(\bm{r}, \hat{p}, \epsilon)
  =
  \begin{bmatrix}
    ig(\bm{r}, \hat{p}, \epsilon) &
    f(\bm{r}, \hat{p}, \epsilon) \\
    \widetilde{f}(\bm{r}, \hat{p}, \epsilon) &
    -i\widetilde{g}(\bm{r}, \hat{p}, \epsilon) \\
  \end{bmatrix}.
\end{equation}
From Eq.\ \eqref{eq:symrel-g-2}, the spin-space matrices $g$ and $f$
are found to have the symmetry
\begin{align}
  &g(\bm{r}, \hat{p}, \epsilon) = -g(\bm{r}, \hat{p}, \epsilon^*)^\dag, \\
  &f(\bm{r}, \hat{p}, \epsilon)
  = -\widetilde{f}(\bm{r}, \hat{p}, \epsilon^*)^\dag
  = -f(\bm{r}, -\hat{p}, -\epsilon)^T,
\end{align}
where the superscript $T$ denotes matrix transpose.

Introducing a spin-space matrix $\mathcal{D}(\bm{r}, \hat{p},
\epsilon)$ called the coherence function \cite{EschrigPRB}, one can
parameterize $\hat{g}(\bm{r}, \hat{p}, \epsilon)$ in a
form that automatically satisfies the normalization condition
$\hat{g}^2(\bm{r}, \hat{p}, \epsilon) = -1$:
\begin{equation}
  \hat{g} + i
  = 2i
  \begin{bmatrix}
    1 \\ -i\tD
  \end{bmatrix}
  \frac{1}{1 - \cD\tD}
  \begin{bmatrix}
    1 & -i\cD
  \end{bmatrix},
  \label{eq:g+i-D}
\end{equation}
or, equivalently,
\begin{equation}
  \hat{g} - i
  = -2i
  \begin{bmatrix}
    i\cD \\ 1
  \end{bmatrix}
  \frac{1}{1 - \tD\cD}
  \begin{bmatrix}
    i\tD & 1
  \end{bmatrix}.
  \label{eq:g-i-D}
\end{equation}
The symmetry relation $\eqref{eq:symrel-g-2}$ implies that the
coherence function has the symmetry
\begin{equation}
  \cD(\bm{r}, \hat{p}, \epsilon^*)^\dag = \cD(\bm{r}, \hat{p}, \epsilon)^{-1}.
  \label{eq:symrel-D}
\end{equation}
Under this parameterization method, the spatial evolution of
$\hat{g}(\bm{r}, \hat{p}, \epsilon)$ is described by the Riccati-type
differential equation \eqref{eq:RCTeq} for $\cD(\bm{r}, \hat{p},
\epsilon)$, instead of the transport-like equation \eqref{eq:Eileneq}
supplemented by the normalization condition.

\section{Random $S$-matrix theory}
\label{appendix:B}

\begin{figure}[t]
  \includegraphics[scale=0.6]{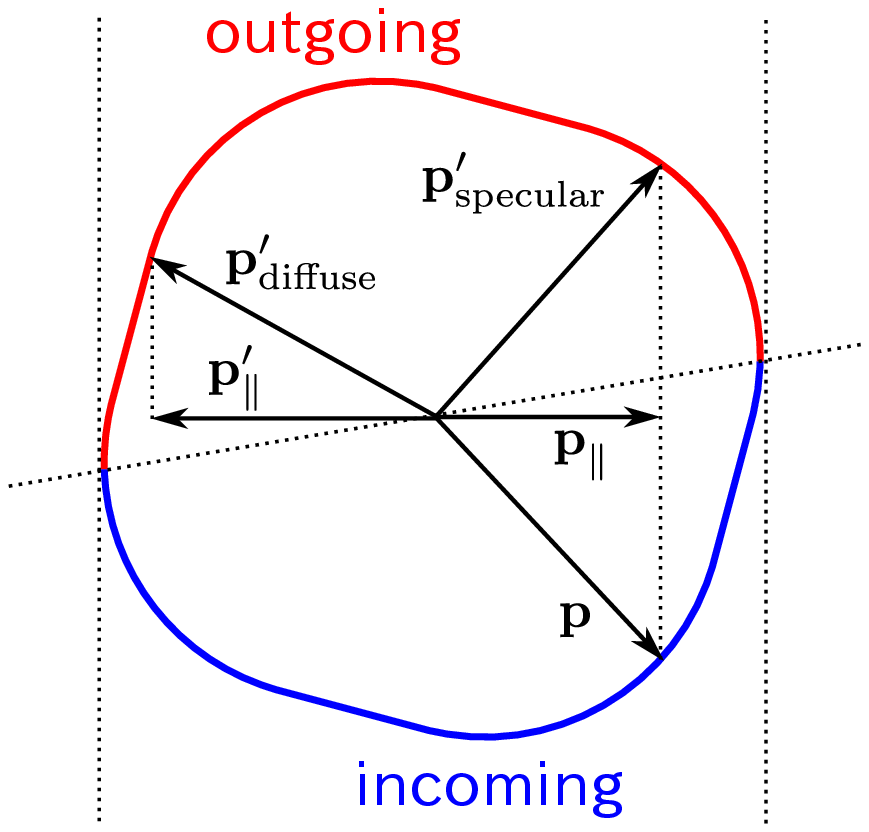}
  \caption{\label{fig:FS} Fermi momenta of the incoming ($\bm{p}$) and
    outgoing ($\bm{p}'$) states.  The incoming (outgoing) state has a
    Fermi velocity towards (away from) the surface.  The Fermi
    velocity is directed outward normal to the Fermi surface.}
\end{figure}

In the random $S$-matrix (RSM) theory \cite{NagatoJLTP}, the surface
effect is incorporated into the quasiclassical theory by introducing
an $S$-matrix $S_{\bm{p}_\|'\bm{p}_\|}$ in the normal state at the
Fermi level and parameterizing it as
\begin{equation}
  S_{\bm{p}_\|'\bm{p}_\|}
  = - \left( \frac{1 - i\eta}{1 + i\eta} \right)_{\bm{p}_\|'\bm{p}_\|}.
  \label{eq:S-def}
\end{equation}
Here, $\bm{p}$ and $\bm{p}'$ are the Fermi momenta of incoming and
outgoing states, respectively, and the subscript $\|$ denotes the
vector component parallel to the surface (Fig.\ \ref{fig:FS}).  The
momentum-space matrix $\eta$ is required to be an Hermite matrix so
that the unitarity of $S$ is assured.  When $\eta = 0$, Eq.\
\eqref{eq:S-def} is reduced to $S_{\bm{p}_\|'\bm{p}_\|} =
-\delta_{\bm{p}_\|'\bm{p}_\|}$. This form of the $S$-matrix
corresponds to the specular surface case, where $\bm{p}_\|$ is
conserved during surface scattering processes.  The diffuse scattering
effect is therefore described by $\eta$. In the RSM theory, every
element of $\eta$ is treated as a random variable to describe the
statistical property of the surface and the statistical average of
$\hat{g}$ is evaluated by employing the self-consistent Born
approximation. A consequence of this procedure is that the diffuse
scattering effect is characterized by the average
$\overline{|\eta_{\bm{p}_\|'\bm{p}_\|}|^2} \equiv \eta^{(2)}(\bm{p}_\|
- \bm{p}_\|')$.

Under this model for the $S$-matrix, the boundary condition for the
averaged Green's function is obtained as
\begin{equation}
  \hat{g}_{\rm out}(\bm{p}_\|, \epsilon)
  = \frac{1 + i\hat{\gamma}_{\bm{p}_\|}(\epsilon)}
  {1 - i\hat{\gamma}_{\bm{p}_\|}(\epsilon)}
  \hat{g}_{\rm in}(\bm{p}_\|, \epsilon)
  \frac{1 - i\hat{\gamma}_{\bm{p}_\|}(\epsilon)}
  {1 + i\hat{\gamma}_{\bm{p}_\|}(\epsilon)},
  \label{eq:bcon-g-RSM}
\end{equation}
where
\begin{align}
  \hat{\gamma}_{\bm{p}_\|}(\epsilon)
  &= \sum_{\bm{p}_\|'} \eta^{(2)}(\bm{p}_\| - \bm{p}_\|')
  \hat{G}_{\bm{p}_\|'}(\epsilon),
  \label{eq:gamma-Gsurf}
  \\
  \hat{G}_{\bm{p}_\|}(\epsilon) \notag
  &= \frac{1}{1 - i\hat{\gamma}_{\bm{p}_\|}(\epsilon)}
  \left[
    \hat{g}_{\rm in}(\bm{p}_\|, \epsilon)
    - \hat{\gamma}_{\bm{p}_\|}(\epsilon)
  \right]
  \frac{1}{1 + i\hat{\gamma}_{\bm{p}_\|}(\epsilon)}
  \notag\\
  &= \frac{1}{1 + i\hat{\gamma}_{\bm{p}_\|}(\epsilon)}
  \left[
    \hat{g}_{\rm out}(\bm{p}_\|, \epsilon)
    - \hat{\gamma}_{\bm{p}_\|}(\epsilon)
  \right]
  \frac{1}{1 - i\hat{\gamma}_{\bm{p}_\|}(\epsilon)}.
  \notag
\end{align}
In Eq.\ \eqref{eq:bcon-g-RSM}, $\hat{g}_{{\rm in} ({\rm out})}
(\bm{p}_\|, \epsilon)$ stands for the surface value of
$\hat{g}(\bm{r}, \hat{p}, \epsilon)$ at the incoming (outgoing) Fermi
momentum with a given parallel component $\bm{p}_\|$. Equation
\eqref{eq:bcon-g-RSM} with $\hat{\gamma}_{\bm{p}_\|}(\epsilon) = 0$
($\eta^{(2)} = 0)$ gives the specular surface boundary condition
\begin{equation}
  \hat{g}_{\rm out}(\bm{p}_\|, \epsilon)
  = \hat{g}_{\rm in}(\bm{p}_\|, \epsilon),
\end{equation}
which means that the quasiclassical propagator is continuous on the
trajectory along a specular reflection process.  This property is lost
at a rough surface because of a finite
$\hat{\gamma}_{\bm{p}_\|}(\epsilon)$.  The Nambu-space matrix
$\hat{\gamma}_{\bm{p}_\|}(\epsilon)$ has symmetries similar to Eqs.\
\eqref{eq:symrel-g-1} and \eqref{eq:symrel-g-2} for the quasiclassical
Green's function, i.e.,
\begin{align}
  \hat{\gamma}_{\bm{p}_\|}(\epsilon)
  &= \hat{\rho}_1\widetilde{\hat{\gamma}}_{\bm{p}_\|}(\epsilon)\hat{\rho}_1
  \label{eq:symrel-gamma-1}
  \\
  &= \hat{\rho}_3\hat{\gamma}_{\bm{p}_\|}(\epsilon^*)^\dag\hat{\rho}_3.
  \label{eq:symrel-gamma-2}
\end{align}

Equation \eqref{eq:bcon-g-RSM} can be rewritten in the form
\begin{align}
  &\hat{g}_{\rm out}(\bm{p}_\|, \epsilon)
  - \hat{g}_{\rm in}(\bm{p}_\|, \epsilon)
  \notag \\ 
  &= 2i\sum_{\bm{p}_\|'} \eta^{(2)}(\bm{p}_\| - \bm{p}_\|')
  [\hat{G}_{\bm{p}_\|'}(\epsilon),\ \hat{G}_{\bm{p}_\|}(\epsilon)].
\end{align}
From this, we readily find
\begin{align}
  0 &= \sum_{\bm{p}_\|}
  \left[
    \hat{g}_{\rm out}(\bm{p}_\|, \epsilon)
    - \hat{g}_{\rm in}(\bm{p}_\|, \epsilon)
  \right]
  \label{eq:flux-conservation} \\
  &\propto
  \int_{\rm out}
  \frac{d^2p_F}{|\bm{v}_{\hat{p}}|}\,
  |v_{\hat{p}}^\perp|\hat{g}_{\rm out}(\bm{p}_\|, \epsilon)
  - \int_{\rm in}
  \frac{d^2p_F}{|\bm{v}_{\hat{p}}|}\,
  |v_{\hat{p}}^\perp|\hat{g}_{\rm in}(\bm{p}_\|, \epsilon),
  \notag
\end{align}
where $v_{\hat{p}}^\perp$ is the Fermi velocity component
perpendicular to the surface. Equation \eqref{eq:flux-conservation}
guarantees that there is no net current across the surface.

The boundary condition for the coherence function is given as
\begin{equation}
  \begin{bmatrix} i\cD_{\rm out}(\bm{p}_\|, \epsilon) \\ 1 \end{bmatrix}
  C_{2 \times 2}
  =
  \frac{1 + i\hat{\gamma}_{\bm{p}_\|}(\epsilon)}
  {1 - i\hat{\gamma}_{\bm{p}_\|}(\epsilon)}
  \begin{bmatrix} i\cD_{\rm in}(\bm{p}_\|, \epsilon) \\ 1 \end{bmatrix},
  \label{eq:bcon-cD}
\end{equation}
where $C_{2 \times 2}$ is an arbitrary spin-space matrix.  The
equivalence between the boundary conditions \eqref{eq:bcon-g-RSM} and
\eqref{eq:bcon-cD} can be confirmed in the following way.  Using the
symmetry relations \eqref{eq:symrel-D}, \eqref{eq:symrel-gamma-1}, and
\eqref{eq:symrel-gamma-2}, one can convert Eq.\ \eqref{eq:bcon-cD}
into the form
\begin{equation*}
  C_{2 \times 2}'
  \begin{bmatrix} i\tD_{\rm in}(\bm{p}_\|, \epsilon) & 1 \end{bmatrix}
  =
  \begin{bmatrix} i\tD_{\rm out}(\bm{p}_\|, \epsilon) & 1 \end{bmatrix}
  \frac{1 + i\hat{\gamma}_{\bm{p}_\|}(\epsilon)}
  {1 - i\hat{\gamma}_{\bm{p}_\|}(\epsilon)},
\end{equation*}
where $C_{2 \times 2}'$ is again an arbitrary spin-space matrix.
Substituting the above two relations for the coherence function into
Eq.\ \eqref{eq:g-i-D}, we obtain Eq.\ \eqref{eq:bcon-g-RSM}.

In the RSM theory, the nature of the boundary condition is specified
by $\eta^{(2)}(\bm{p}_\|-\bm{p}_\|')$.  We can describe the surface
effect from the specular to the diffusive limit (Fig.\
\ref{fig:rough_surface}) in a unified way by expressing it as
\begin{equation}
  \eta^{(2)} = \frac{2W}{\sum_{\bm{p}_\|}1}, \ \
  W = \frac{1 - \sqrt{R}}{\left(1 + \sqrt{R} \right)^2},
  \label{eq:eta2-W}
\end{equation}
where $R$ is a momentum-independent parameter. Physically, $R$
corresponds to the surface specularity,
\cite{SeijiJPSJ2015,MurakawaPRL,MurakawaJPSJ,Okuda} which is defined
as the specular reflection probability in the normal state at the
Fermi level. In fact, evaluating the statistical average of
$|S_{\bm{p}_\|'\bm{p}_\|}|^2$ with Eq.\ \eqref{eq:eta2-W}, we obtain
\cite{Yamamoto}
\begin{equation}
  \overline{|S_{\bm{p}_\|'\bm{p}_\|}|^2}
  = R \delta_{\bm{p}_\|'\bm{p}_\|} + \frac{1 - R}{\sum_{\bm{p}_\|}1}.
  \label{eq:S2-R}
\end{equation}
It is obvious that the specular surface corresponds to $R = 1$. The
diffuse limit, where surface scattering occurs in any possible
direction with equal probability $1/\sum_{\bm{p}_\|}1$, is achieved
for $R = 0$.  It follows that the above one-parameter model for
$\eta^{(2)}$ provides a simple interpolation formula connecting the
specular and diffuse limits.

When the boundary condition is parameterized with Eq.\
\eqref{eq:eta2-W}, $\hat{\gamma}_{\bm{p}_\|}(\epsilon)$ is independent
of $\bm{p}_\|$ and is given by
\begin{align}
  \hat{\gamma}(\epsilon)
  &= \frac{2W}{1 + 2W + \hat{\gamma}^2(\epsilon)}\,\hat{g}_0(\epsilon),
  \label{eq:gamma-g0}
  \\
  \hat{g}_0(\epsilon)
  &= \langle \hat{g}_{\rm in}(\bm{p}_\|, \epsilon) \rangle_{\bm{p}_\|}
  = \langle \hat{g}_{\rm out}(\bm{p}_\|, \epsilon) \rangle_{\bm{p}_\|},
\end{align}
where
\begin{equation}
  \langle \cdots \rangle_{\bm{p}_\|}
  = \sum_{\bm{p}_\|} (\cdots) / \sum_{\bm{p}_\|}1.
\end{equation}
Because of the symmetries \eqref{eq:symrel-g-1} and
\eqref{eq:symrel-gamma-1}, $\hat{g}_0(\epsilon)$ and
$\hat{\gamma}(\epsilon)$ have the matrix structures
\begin{align}
  \hat{g}_0(\epsilon)
  &=
  \begin{bmatrix}
    ig_0(\epsilon) & f_0(\epsilon) \\
    \widetilde{f}_0(\epsilon) & -i\widetilde{g}_0(\epsilon)
  \end{bmatrix},
  \label{eq:g0-ex}\\
  \hat{\gamma}(\epsilon)
  &=
  \begin{bmatrix}
    ia(\epsilon) & b(\epsilon) \\
    \widetilde{b}(\epsilon) & -i\widetilde{a}(\epsilon)
  \end{bmatrix}.
  \label{eq:gamma-ex}
\end{align}

Finally, we note that the RSM theory in the diffuse limit gives the
same boundary condition obtained from Ovchinnikov's rough surface
model \cite{Ovchinnikov,VorontsovSauls}. To see this, let us first
assume that the matrix $\hat{\gamma}(\epsilon)$ in the diffuse limit,
which we denote by $\hat{\gamma}_{\rm DL}(\epsilon)$, has the property
\begin{equation}
  \hat{\gamma}_{\rm DL}^2(\epsilon) = -1
  \label{eq:gammaDL2}
\end{equation}
similar to the normalization condition for the quasiclassical Green's
function. It can be shown that Eq.\ \eqref{eq:gammaDL2} is in fact
satisfied in the normal state; the quasiclassical Green's function in
the normal state is given as $\hat{g}_N(\epsilon) = {\rm sgn}({\rm
  Im}[\epsilon])i\hat{\rho}_3$. Then Eq.\ \eqref{eq:gamma-g0} has
the solution
\begin{equation}
  \hat{\gamma}(\epsilon)
  = \frac{1 - \sqrt{R}}{1 + \sqrt{R}}\,\hat{g}_N(\epsilon).
\end{equation}
When $R = 0$, $\hat{\gamma}(\epsilon) = \hat{g}_N(\epsilon)$ and hence
Eq.\ \eqref{eq:gammaDL2} holds.  Assuming that it also holds in SC
states, we can write the boundary condition \eqref{eq:bcon-cD} in the
form
\begin{gather}
  [1 - i\hat{\gamma}_{\rm DL}(\epsilon)]
  \begin{bmatrix}
    i\cD_{\rm out}(\bm{p}_\|, \epsilon) \\ 1
  \end{bmatrix}
  = 0,
  \label{eq:bcon-cD-DL} \\
  \hat{\gamma}_{\rm DL}(\epsilon) = \hat{g}_0(\epsilon).
  \label{eq:gamma-g0-DL}
\end{gather}
Equation \eqref{eq:bcon-cD-DL} tells us that $\cD_{\rm out}(\bm{p}_\|,
\epsilon)$ in the diffuse limit is independent of $\bm{p}_\|$. Noting
this and using Eqs.\ \eqref{eq:g+i-D} and \eqref{eq:g-i-D}, we readily
find that $\hat{g}_0(\epsilon)$ has the property
$\hat{g}_0^2(\epsilon) = -1$, which justifies the assumption of Eq.\
\eqref{eq:gammaDL2}.  From Eqs.\ \eqref{eq:g0-ex},
\eqref{eq:bcon-cD-DL}, and \eqref{eq:gamma-g0-DL}, we obtain
\begin{equation}
  \cD_{\rm out}(\bm{p}_\|, \epsilon)
  = \frac{1}{g_0(\epsilon) + 1}f_0(\epsilon)
  = \frac{1}{\widetilde{f}_0(\epsilon)}[\widetilde{g}_0(\epsilon) - 1].
  \label{eq:bcon-cD-DL-ex}
\end{equation}
The second equality holds because $\hat{g}_0^2(\epsilon) =
-1$. Equation \eqref{eq:bcon-cD-DL-ex} coincides with the boundary
condition derived by Vorontsov and Sauls \cite{VorontsovSauls} using
Ovchinnikov's rough surface model.


\end{document}